\documentclass[aps,prl,twocolumn,superscriptaddress]{revtex4}

\usepackage{graphicx}

\begin{document}

\title{
Low-lying electronic structure of doped triangular cobaltites}

\author{A. P. Kuprin}
\affiliation{Advanced Light Source, Lawrence Berkeley Laboratory,
Berkeley, CA 94305}
\author{D. Qian}
\affiliation{Joseph Henry Laboratories, Department of Physics,
Princeton University, Princeton, NJ 08544}
\author{Y.-D. Chuang}
\affiliation{Advanced Light Source, Lawrence Berkeley Laboratory,
Berkeley, CA 94305}
\author{M. Foo}
\author{R. J. Cava}
\affiliation{Department of Chemistry, Princeton University,
Princeton, NJ 08544, USA.}
\author{M. Z. Hasan}\affiliation{Joseph Henry
Laboratories, Department of Physics, Princeton University,
Princeton, NJ 08544}\affiliation{Princeton Center for Complex
Materials, Princeton University, Princeton, NJ 08544, USA.}

\date{\today}

\begin{abstract}
We report detailed Fermi surface topology and quasiparticle dynamics
of the host cobaltate Na$_{0.7}$CoO$_{2}$. A direct mapping of the
Fermi surface is carried out by angle-resolved photoemission
spectroscopy (ARPES). Fermi surface at 16~K is a hole-pocket
centered around the $\Gamma$~-~point. The highly correlated nature
of the electron liquid in this system is evidenced from strong
on-site Coulomb interaction seen through a valence satellite and via
renormalization of the quasiparticle dispersion. The quasiparticle
exhibits spectral coherence only below 150K which includes the
non-Fermi liquid regime.
\end{abstract}



\maketitle

\section{\label{sec:intro}Introduction\protect}
The family of sodium cobalt oxyhydrates or oxides (cobaltites)
Na$_{x}$CoO$_{2}$\ (with x varying from 0.7 to 0.3) attracted a
great deal of attention recently after the discovery of
superconductivity around 5~K in the x$\sim$0.3 compound
\cite{takada03}. Though there are several clear differences in
crystal structure and electronic structure between cobaltites and
high~-~\textit{{T$_{\rm c}$}}\ cuprates both systems have layered
structure with electronic two-dimensionality, and typical strong
electron correlations of a doped Mott insulator. Moreover, the
triangular lattice of cobaltites induces spin frustration and hence
may lead to Anderson's RVB state \cite{anderson73} that can also
result in a non-s-wave order parameter \cite{wang03cm}. Low
temperature ferromagnetic fluctuations observed in cobaltites may
lead to a p~-~wave superconducting state\cite{singh03}. In addition,
the cobaltite with $x\,=\,0.7$, being the host compound for the
family of lower $x$, demonstrates its own unusual physical
properties like linear rise of resistivity with temperature up to
100~K,\cite{wang03nat} no saturation of the anomalous Hall signal up
to 500~K,and extraordinary high and magnetic field dependent
thermopower\cite{wang03nat, terasaki97}. Hence, the details of
complete Fermi surface topology and of quasiparticle dynamics are
highly desirable for understanding the unconventional physics of
this promising class of materials. This work reports such direct
measurements performed by the angle-resolved photoemission
spectroscopy (ARPES) on the host compound Na$_{0.7}$CoO$_{2}$.

Single crystals of Na$_{0.7}$CoO$_{2}$\ prepared by the flux method
\cite{wang03nat} were kept under dry nitrogen gas and screened by
x~-~ray diffraction in Laue geometry. ARPES experiments were carried
out with Scienta analyzers at the Advanced Light Source Beamlines
7.0.1 and 12.0.1. Photon energy covered 30~eV - 90~eV range with
energy resolution better than 30~meV. The complete Fermi surface
topology was measured in the image mode of the programmable
three-axis goniometer at the "Electronic structure Factory"
endstation of Beamline 7.0.1. Fermi surface areas close to principal
symmetry points were also obtained independently from a set of
energy spectra for different angles of incidence and emission.
Multi-angle mode allowed acquisition of energy spectra in
$9.5^{\circ}$ wide emission angle window with an angular resolution
better than 1$\%$ of the Brillouin zone (BZ). To obtain clean and
flat surface, samples were cleaved under ultra-high vacuum of
$5\times10^{-11}$~torr or better at 16~K.

\begin{figure}[b]
\includegraphics[width=9cm]{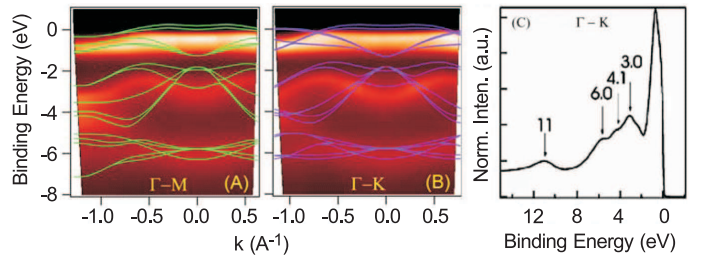} \caption{\label{fig:VB}
Valence band of Na$_{0.7}$CoO$_{2}$: (a) and (b) intensity plots
along $\Gamma$~-~M and $\Gamma$~-~K lines respectively; both images
obtained with 65~eV photons, curves are the LDA results
from,\cite{singh00}  (c) averaged energy dispersion curve (EDC) for
$\Gamma$~-~K orientation obtained with 90~eV photons showing
correlation satellite at around 11~eV. }
\end{figure}

\section{\label{sec:elstruct}Electronic structure\protect}

The valence band dispersions along $\Gamma$~-~K and $\Gamma$~-~M
principal symmetry lines follow the general trends of LDA
calculations \cite{singh00} including energy positions, widths, and
dispersion of recognized features at binding energies around 0.7~eV,
3.0~eV, 4.1~eV, 6.0~eV, and the width of the density of states of
about 7~eV as shown in Fig-1(a) and Fig-1(b) respectively. Some
bands may be less visible due to photon energy and polarization
dependence of the matrix elements\cite{damascelli03}. A well defined
peak is also observed at higher binding energies around 11~eV as
show in Fig-1(c). Its resonant enhancement near 3p $\to$ 3d
excitation together with the photon energy dependence of the
Fano~-~type interference are shown in Fig-2 and allow to identify it
as a correlation satellite that arises from strong on-site Coulomb
interaction and is consistent with cluster calculations with strong
Hubbard~-~{\it U}($\sim$5eV), providing strong evidence for the
electron's highly correlated nature.

\begin{figure}[t]
\includegraphics[width=8cm]{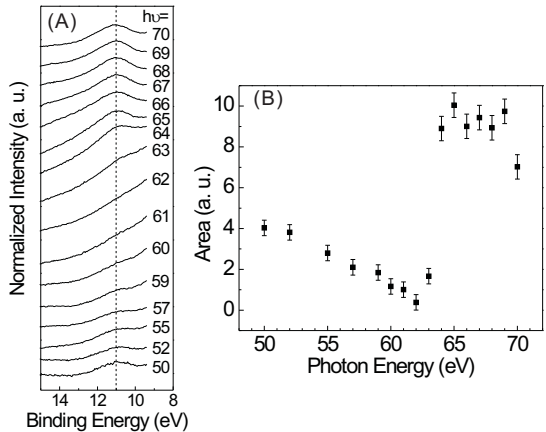} \caption{\label{fig:res}
Resonant enhancement of the correlation satellite at 11~eV: (a) EDCs
taken at different photon energies around Co 3p $\to$\ 3d transition
at $\sim\,63$~eV, and shifted for clarity. (b) Background subtracted
integrated intensity of the 11~eV feature shows Fano-type
interference resonance.}
\end{figure}

\begin{figure}[t]
\includegraphics[width=7cm]{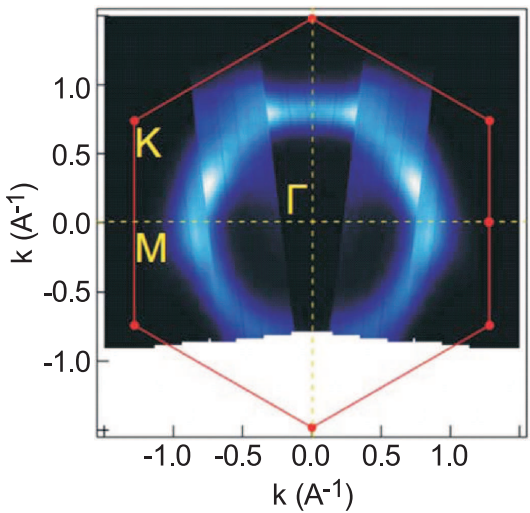} \caption{\label{fig:FS}
The energy integrated k-resolved intensity map. Fermi surface of
Na$_{0.7}$CoO$_{2}$\ is the inner edge of the k-resolved intensity
map (a hexagonal hole pocket).}
\end{figure}

Quasiparticle features were found near the Fermi crossing along all
direction from zone boundary towards $\Gamma$ point. By tracing the
crossing point in the whole Brillouin zone, we get the Fermi
surface. The shape of the Fermi surface at 16~K is established to be
an anisotropic hexagonal-like hole-pocket centered around the
$\Gamma$~-~point as shown in Fig-3 \cite{hasan04} in agreement with
LDA calculations\cite{singh00} by integrating spectral weight from
75~meV below Fermi level (E$_F$) to 25~meV above it with 90~eV
photons. In disagreement with these calculations, no small satellite
pockets are observed around the central one. This may reflect strong
electron correlations that may drive the minority bands out of
E$_F$. The size of the central pocket is
$0.65\,\pm\,0.1~$\AA$^{-1}$. Intensity plot and energy dispersive
curves (EDCs) near E$_F$ crossing along $\Gamma$~-~M and
$\Gamma$~-~K symmetry lines are presented in Fig-4. These images
were collected from the very same sample during the same
experimental run and conditions following only $30^{\circ}$
azimuthally rotation. The quasiparticle feature strongly depends on
the momentum ({\bf k}). Its spectral weight being well defined only
in the limited area of the BZ may be explained by drastic lifetime
shortening on departure from Fermi level similar to the case of
cuprates and other correlated systems. For both principal
orientations this quasiparticle feature disperses only slightly,
overall hardly more than 100~meV. Thus, it is at least five times
narrower than for the cuprates which is an order of magnitude
renormalization from the mean field calculations
\cite{wang03cm,singh00} and hence may be responsible for an order of
magnitude enhancement of thermopower \cite{terasaki97} and
electronic specific heat \cite{ando99}. Along both principal
symmetry lines it crosses Fermi surface in the same direction - from
the BZ boundary toward $\Gamma$~-~point, giving the negative sign of
the nearest neighbor single particle hopping {\it t}. The
quasiparticle (occupied) bandwidth is about $100\,\pm\,30$~meV. The
value of {\it t} can be estimated from tight-binding approximation
and for the lattice with hexagonal symmetry (bandwidth =
9\textit{t}) leads to $11\,\pm\,3$~meV. Such small values imply a
very small fermion degeneracy temperature in comparison with normal
metals. It is worth mentioning that these values are of the same
order as exchange coupling $J\,\sim\,10$~meV for the compounds of
this family \cite{wang03nat} and hence, charge motion might be
strongly perturbed by the spin fluctuations. Na$_{0.7}$CoO$_2$ has
small Fermi velocity less than $\sim$ 0.3 eV$\cdot$\AA, which can be
estimated from EDC and momentum distribution curve (MDC) as well.
This value is almost 6 times smaller than the nodal Fermi velocity
for the cuprates of 1.8~eV$\cdot$\AA, which is in agreement with
much wider dispersion of the cuprate quasiparticle bands
\cite{damascelli03}.

\begin{figure}
\includegraphics[width=6cm]{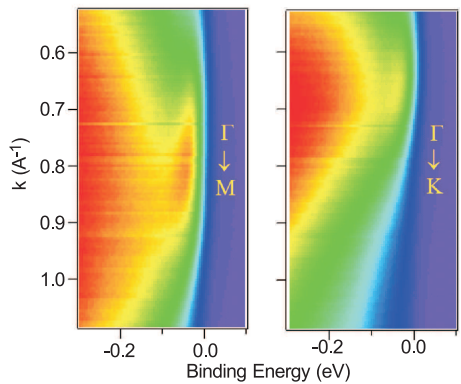} \caption{\label{fig:qpGM}
Quasiparticle feature near E$_F$ along $\Gamma$~-~M and $\Gamma$~-~K
cuts.}
\end{figure}

\section{\label{sec:Tdep}Temperature dependence of quasiparticle\protect}

We performed temperature dependent measurements of the spectral
weight near E$_F$ presented in Fig-5. On raising temperature
quasiparticle weight steadily decreases up to about 120~K, where it
becomes indistinguishable from the background. Extrapolation to zero
places the temperature value to around 150K. Similar behavior is
observed by cooling the samples. This temperature roughly matches
the crossover region where resistivity changes its temperature
dependence from linear at lower temperatures to a much steeper one
shown in Fig-5(c) \cite{wang03nat}. Quasiparticle transport property
may be connected to the inherently frustrated nature of local
antiferromagnetic interactions in a triangular lattice as the
strength of the exchange interactions {\it J} and single-particle
hopping {\it t} are of the same order. It is interesting to note
that this temperature scale is also relevant to peculiarities of
Hall coefficient and thermopower \cite{wang03nat}. Such a low energy
scale present in a variety of physical properties of this compound
still lacks proper theoretical explanation. It is desirable to carry
out inelastic x-ray scattering to study the collective charge
dynamics in this system which would be complementary to the ARPES
study \cite{hasan}.

\begin{figure}
\includegraphics[width=9cm]{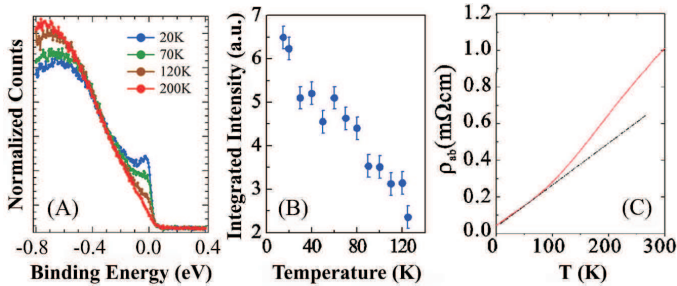} \caption{\label{fig:temp}
Temperature scale for Na$_{0.7}$CoO$_{2}$: (a) T-dependence of
quasiparticle weight along  $\Gamma$~-~M line, (b) integrated
spectral weight after background subtraction, and (c) in-plane
resistivity. }
\end{figure}

\section{\label{sec:concl}Conclusions\protect}

In summary, we observed a number of electronic features that
indicate the existence of strong Coulomb correlations in
Na$_{0.7}$CoO$_{2}$: intense valence band satellites, a highly
renormalized quasiparticle band near Fermi level and the existence
of quasiparticle coherence only below 150K. The small low energy
scales for the effective single-particle hopping {\it t}, magnetic
exchange coupling {\it J} and non-Fermi liquid behavior observed
opens the way for unconventional physics in the cobaltates.

\begin{acknowledgments}
\label{sec:ack}

We thank A. Fedorov, E. Rotenberg, R. Kimmerling, K. Rossnagel, H.
Koh, J. Denlinger, and B. Mesler for technical support, A.K.
acknowledges financial support from ALS, and M.Z.H. partial support
through the NSF~-~MRSEC (DMR~-~0213706) grant and characterization
supported by DOE grant DE-FG02-05ER46200. Materials synthesis
supported by DMR-0213706 and the DOE, Grant No.
DE~-~FG02~-~98~-~ER45706.

\textit{Note added} : Subsequent to our work H.B. Yang
\textit{et.al.} \cite{yang}, reported a new measurement of FS on a
sample with similar doping-level as ours. In this study, only one
Fermi surface is found which is reported to be of size K$_f$ $\sim$
$0.6\,~$\AA$^{-1}$ \cite{yang}. This is in agreement with our
reported value within the error bar.

\end{acknowledgments}

\end{document}